\documentclass[11pt]{article} 
\usepackage{epsfig}
\usepackage{amssymb}
\usepackage{amsmath}
\usepackage{mathpple}

\newcommand{\ph} {\mbox{$\xi^2_j$}}

\begin{document}

\title{Photon Folding for Imaging in Non-Focusing Telescopes}
\author{JaeSub Hong \\ \\
\it Columbia Astrophysics Laboratory, Columbia University \\
\it 550 West 120th Str, New York, NY 10027, USA \\
\it flame@zen.phys.columbia.edu}
\date{}

\maketitle

\begin{abstract}We present a new technique -- {\it photon folding} -- for
imaging in non-focusing telescopes.  Motivated by the epoch-folding
method in timing analysis, the photon folding technique directly
provides the statistical significance of signals in images, using
projection matrices.  The technique is very robust against common
imaging problems such as aspect errors and non-uniform background.  The
technique provides a deterministic recursive algorithm for improving
images, which can be implemented on-line.  Higher-order photon
folding technique allows a systematic correction for coding noises,
which is suitable for studying weak sources in the presence of highly
variable strong sources.  The technique can be applied to various types
of non-focusing telescopes such as coded-aperture optics, rotational
collimators or Fourier grid telescopes.

keywords: coded-aperture system, modulation collimator, Fourier grid
telescope, epoch-folding method
\end{abstract}

\section{Introduction}

Localization or image reconstruction of astrophysical sources at hard
X-ray/$\gamma$-ray energies by non-focusing telescopes, such as a
coded-aperture system or modulation collimator, has always been a
challenging topic. This is largely due to an intrinsically low
signal-to-noise ratio (SNR) in non-focusing systems compared to focusing
instruments \cite{Fenimore:78}.  Consequently, the next generation hard
X-ray telescopes, such as {\it HEFT}\footnote{High Energy Focusing
Telescope ({\it HEFT}) is a balloon-borne experiment that will use
depth-graded multilayer  optics and Cadmium Zinc Telluride pixel
detectors to image astrophysical    sources in the hard X-ray (20 -- 100
keV) band.} or {\it Constellation-X}/HXT\footnote{The {\it
Constellation-X} Observatory is a next-generation X-ray telescope
satellite, planned as four individual X-ray space telescopes operating
together. The Hard X-ray Telescope (HXT) on {\it Constellation-X} is designed
to image X-rays from 5 to 100 keV with resolution better than 1$'$ over
a 8$'$ field of view.  X-rays are collected by graded multilayer
reflective      optics and imaged with a position sensitive X-ray
detector.}, push focusing optics up to energies as high as 100 keV
\cite{Heft:97,constX:00}. In order to study astrophysical sources at
energies above $\sim$ 100 keV, however, we still have to rely on some
sort of non-focusing schemes for the time being
\cite{exist:99,BeppoSax:99}.

For example, gamma-ray bursts, one of the long-standing mysteries in
astronomy, usually have about 90\% of their measured flux in the energy
range from $\sim$ 100 keV to a few MeV.  Their secret is expected to be
solved or at least resolved greatly by non-focusing instruments such as
{\it Swift}\footnote{The {\it Swift} is a gamma-ray burst explorer, which will
carry two X-ray  telescopes and one UV/optical telescope to enable the
most detailed observations of gamma-ray bursts to date.} in the near
future \cite{Swift:99}.  The fast, accurate localization of bursts by
non-focusing optics is required to guarantee effective multi-frequency
follow-up observations, which may be important for understanding the
physics of the bursts, as demonstrated in recent
{\it BeppoSAX}\footnote{{\it BeppoSAX} is an Italian X-ray astronomy
satellite with the wide spectral coverage ranging from 0.1 to 
200 keV.} observations \cite{BeppoSax:99}.

There are many techniques for reconstructing images in non-focusing
telescopes. For example, in coded-aperture systems, two types of methods
are largely used -- correlation and inversion methods.  Both methods are
subject to either coding noise or quantum noise, which leads to spurious
fluctuations in images.  In order to improve the quality of images, a
lot of different techniques are also employed such as pixon or maximum
entropy method (MEM) \cite{Pixon:99,MemPixon:97}.  These methods are
computationally intensive and, quite often, requires extensive control
over somewhat uncertain parameters in order to produce satisfactory
results.

More importantly, in the case of applying non-linear refinement
algorithms like pixon or MEM on images reconstructed by correlation or
inversion methods, time-independent point spread functions (maybe
position dependent) are usually used.  Therefore, without prior
knowledge of the sources intensity change, it is rather difficult to
deal with variable sources when the sources in the FOV have apparent
movements due to motions of telescope relative to the sky.

Here we introduce a new method -- {\it photon folding} -- for imaging
in non-focusing optics.  The basic idea of photon folding is motivated
by the well-known epoch-folding method in timing analysis, which is a
powerful, yet very simple procedure for finding periodicities in data
\cite{epf:83, epf:96}.  The photon-folding method provides a natural way to
assess the statistical significances of images, which is inherited from
the epoch-folding method.  Our technique uses only projection matrices
and can be applied to various types of non-focusing optics.  The
technique is flexible enough to include corrections for aspect errors
and non-uniform background.

The criterion given by the photon folding technique allows a recursive
method for improving images without human supervision of any
parameters, and thus it can be implemented as on-line data
processing.  In practice, the relative orientation between telescopes
and astrophysical sources continues to change during observations and
the intensity of sources may also vary. In such cases, recursive
methods cannot always effectively extract the location or intensity of
a certain source particularly in the presence of highly variable
stronger sources.  High-order photon folding provides a systematic
correction for coding noises in such cases beyond what conventional
methods can provide.

We demonstrate the idea using a two-dimensional coded-aperture system.
The technique can be readily generalized to other systems such as
modulation collimators.

\section{Photon Folding Theory}

\subsection{Basic Concept}

Consider a typical coded-aperture telescope with a uniform background,
where projection matrix $P_{ij}$ can have either 0 (closed) or 1 (open
mask element).  A source $s_j$ from direction $j$ in the sky can
generate counts $h_i$ at bin $i$ in the detector as
 \begin{align}
	h_i & = \sum_j P_{ij} s_j + b_i,
\end{align}
where $b_i$ represents the background count at bin $i$ in the detector
($H=\Sigma h_i$). We use index symbol $i$ for detector bins and all the other
index symbols ($j$,$k$,$l$) represent sky coordinates.
Usually the expected image $\hat{s}_j$
is reconstructed as 
 \begin{align}
	\hat{s}_j & = \sum_i G_{ji} h_i,
\end{align}
where $G_{ji}$ is a reconstruction matrix, usually given by 
inversion or correlation methods.

The epoch-folding method in timing analysis searches for any deviation
in the data from a flat distribution.  Like the $\chi^2$-test
in the epoch-folding method, we can define $\xi^2$ as
	\begin{align}
		\xi^2 &= \sum_i \frac{ \left( h_i - \bar{h} \right)^2}
			{ \bar{h}}, 
	\end{align} where $\bar{h}$ is the average of $h_i$.  
In a typical  coded-aperture system whose overall transparency is
close to 0.5, $\xi^2$ from a point source has the largest value
while $\xi^2$ from a flat field observation is close to a
minimum.  Consequently, the above $\xi^2$ indicates the presence of sources
within the field of view (FOV), but it cannot reveal the location of
the sources.

In general, the fraction ($\rho$) of open mask elements seen by the
detector depends on the direction in the sky, i.e.  
	\begin{align}
	\rho\equiv\rho_{j} = 
		\sum_{i:P_{ij}=1} \frac{1}{M} 
		= \sum_i \frac{P_{ij}}{M}, 
	\end{align} where $M$ is the number of detector pixels.
We can also define $\alpha_j$, the fraction of detected photons which
can originate from direction $j$ in the sky,
	\begin{align}
		\alpha_j &= \sum_{i:P_{ij}=1} \frac{h_i}{H} 
		= \sum_i \frac{h_i P_{ij}}{H}. \label{eq:need}
	\end{align}

\begin{figure} \begin{center}
\epsfig{figure=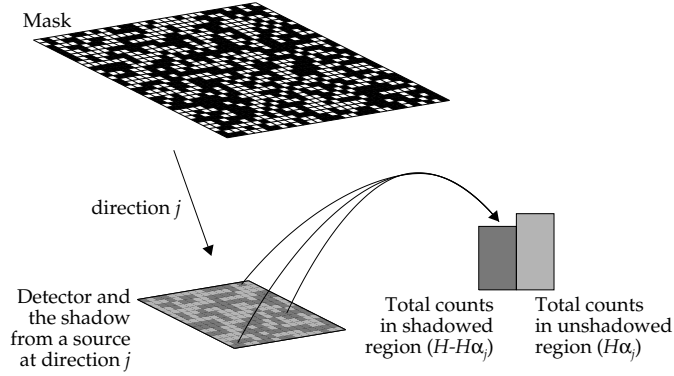,width=10.0cm,clip=} 
\caption{Illustration of photon folding: for a given direction in the
sky, the detected photons are folded into two regions -- shadowed/unshadowed.}
\label{fig:illust}
\end{center} \end{figure}

For a given direction in the sky at any moment, one can divide the
detector space into two separate regions; the region that can have
source photons from the given direction and the other region that is
shadowed by the mask (Fig.~\ref{fig:illust}).  If there is no source
within the FOV, the ratio of the total counts in these two regions will
be simply the ratio of open/closed mask elements, i.e.~$\alpha_j \sim
\rho_j$.  If there is an appreciable source located at the given
direction, $\alpha_j > \rho_j$.  The difference of $\alpha_j$ and
$\rho_j$ indicates the probability of having a source at the given
direction in the sky.

For the direction $j$ in the sky, by folding detected photons into the
two regions (Fig.~\ref{fig:illust}), we define \ph\ of the photon
folding like the $\chi^2$-test in the epoch folding method as
 \begin{align}
	\ph\  &= \frac{ \left( H \alpha_j - H \rho_j \right)^2}
			{H \rho_j} +\frac{ \big( H (1-\alpha_j) - H (1-\rho_j)
			\big)^2} {H (1-\rho_j)}  \nonumber \\
 	  &= H\frac{\left( \alpha_j - \rho_j \right)^2}{(1-\rho_j)\rho_j}.
			\label{eq:def} 
 \end{align} 
In the absence of sources, if $H \alpha_j$ and $H (1 - \alpha_j)$ are
governed by a normal distribution, the \ph\ of the photon folding
follows $\chi^2_{2-1}$ statistics. When $p_1$ is the probability
density of $\chi^2$ statistics with one degree of freedom, the
probability ($Q_1$) of having $\chi^2$ greater than $\eta$ is given as
 \begin{align}
	Q_1(\chi^2_1\ge\eta) & = \int^{\infty}_{\eta}p_1(\chi^2)d\chi^2,
 \end{align}
and then, the confidence level $C$ is related to $Q_1$ as,
\begin{align}
	1-\frac{C}{100} &= K Q_1(\chi^2_1\ge\eta), \label{eq:confi}
\end{align} where $K$ is the number of sky bins.

If there is only one source ($s_j$) within the FOV, the reconstructed
image ($\hat{s}_j$) can
be directly calculated from $\alpha_j$ as
 \begin{align}
	\hat{s}_j  = \frac{H(\alpha_j - \rho_j)}{M (1-\rho_j)\rho_j} & =
	s_j,\label{eq:image} 
 \end{align} 
The above formula for $\hat{s}_j$ by photon folding is similar to unbiased
balanced correlation \cite{Hammersley:92} and becomes identical to the cross
correlation method \cite{Skinner:95} when $P_{ij}$ can have only 0 or 1
and there is only one time bin.  For the uniformly redundant array (URA)
systems where $\rho_j = \rho$, the above formula reduces to the
balanced correlation method where $G_{ji}$ is given by
 \begin{align}
	G_{ji} = \frac{P_{ij}-\rho}{1-\rho}.  \end{align} This is not
surprising since photon folding searches for a deviation from a
flat background distribution similar to the epoch folding method, and
the balanced correlation is optimized to cancel out a flat background in
the detector.

The immediate advantage of the photon folding is that photon folding
provides the statistical significance of signals in images without
additional calculations of variances. That is, from Eq.~(\ref{eq:def})
and (\ref{eq:image}),
 \begin{align}
	\frac{\ph\ }{M^2 (1-\rho_j)\rho_j}  = & \frac{\hat{s}_j^2}{H},
 \end{align} which is similar to the case of focusing optics where
the strength of signals in images directly indicates the
significance of the signals. Now we derive a generalized photon
folding formula which can account for aspect errors and non-uniform
background.

\subsection{Generalization}

For a given direction ($j$) in the sky and for a given time bin ($t$),
the detector space can be ranked by the effective transparency through 
collimators, mask elements, detector efficiency, etc.  One can assign
this transparency to the projection matrix, which can now have any value
from 0 to 1.  Thus, in general, the projection matrix $P_{ij}$ can be
defined as \begin{align}
	P_{ij}(t) & = 0 \dots \delta \dots 1,
\end{align} where $\delta$ is a boundary threshold between shadowed/unshadowed 
regions. Let $K$, $M$ and $T$ be the total number of bins in sky, detector
and time bin respectively ($T$ could be the total duration of the observation
when the duration of time bins varies).

Now, we assume that the detector background is known to follow a pattern
$\bar{b}_i(t)$, which can be estimated by flat-field observations or
on-source observation of weak sources.  The detected counts $h_i(t)$
are given as
\begin{align}
	h_i(t) & = \sum_j P_{ij}(t) s_j(t) +b_i(t), \nonumber \\
 	H(t) & = \sum_i h_i(t),
\end{align} where $b_i(t)$ is the actual background counts at
detector bin $i$ and at time bin~$t$.

For an arbitrary quantity $x_i(t)$ in time bin
$t$ and detector bin $i$, we define the following quantities. 
\begin{align}
	X(t) & = \sum_i x_i(t), \nonumber \\ \lambda_{j_1\dots j_n:l_1\dots
	l_n}[x_i(t),t]
		&=\frac{1}{ X(t) } \sum_{i : P_{ij_1}(t) \ge \delta,
		\atop \dots P_{ij_n}(t) \ge \delta} x_i(t) P_{il_1}(t)
		\cdots P_{il_n}(t),\nonumber \\ \lambda_{j'_1\dots j_n:l_1 \dots
	l_n}[x_i(t),t]
		&=\frac{1}{X(t) } \sum_{i : P_{ij_1}(t) < \delta, \atop
		\dots P_{ij_n}(t) \ge \delta} x_i(t) P_{il_1}(t) \cdots
		P_{il_n}(t),
\end{align} where the prime ($'$) notation is used to represent a
complementary region. For example, \begin{align}
	\lambda_{jk:l}[x_i(t),t] + \lambda_{j'k:l}[x_i(t),t] & =
	\lambda_{k:l}[x_i(t),t].  \end{align} 
We introduce $\lambda$ to simplify the following definitions of $\alpha$,
$\beta$, and $\rho$ and their explicit definitions are
given in the appendix.  In terms of $\lambda$, we define
\begin{align}
	\alpha_{j_1\dots j_n}(t)
		&=\frac{\displaystyle \lambda_{j_1\dots
		j_n:}[h_i(t),t]}
			{\lambda_{j_1\dots j_{n-1}:}[h_i(t),t]}, \nonumber
	 \\ \beta_{j_1\dots j_n}(t)
		&=\frac{\lambda_{j_1\dots j_n:}[\bar{b}_i(t),t]}
		{\lambda_{j_1\dots j_{n-1}:}[\bar{b}_i(t),t]}, \nonumber \\
	\rho_{j_1\dots j_n:l_1\dots l_n}(t)
		&=\frac{\lambda_{j_1\dots j_n:l_1\dots l_n}[1,t]}
		{\lambda_{j_1\dots j_{n-1}:}[1,t]}.
\end{align} The above notation allows an intuitive description of the
complicated calculations. It should be noted that here
$\beta_{...}(t)$ and $\rho_{...}(t)$ can be pre-programmed or
pre-calculated.  Again in order to simplify the notation, we introduce the
sum on time bins as
\begin{align}
	\{X \} & = \sum_t X(t).  \end{align}

Then, the \ph\ of photon folding is defined as
\begin{align}
	\xi^2_{j} & = 
		\frac{\displaystyle 
		\big\{ H \alpha_{j} - H \beta_{j}\big\}^2}
		{\displaystyle \big\{ H \beta_{j}\big\}}
		+
		\frac{\displaystyle 
		\big\{ H \alpha_{j'} -  H \beta_{j'}\big\}^2}
		{\displaystyle \big\{ H \beta_{j'}\big\}} \nonumber \\
	& = 
		\frac{\displaystyle \big\{ H\big\} 
			\big\{ H (\alpha_{j}
			- \beta_{j}) \big\}^2}
		{\displaystyle \big\{ H \beta_{j}\big\}
			\big\{ H (1-\beta_{j})\big\}}. \label{eq:exactXI}
\end{align} 

The first-order reconstructed image $\hat{s}_j$ is given as
\begin{align}
	\frac{\{\hat{s}_j\}}{T} & = \frac{\displaystyle
		\big\{ H (\alpha_{j} - \beta_{j}) \big\}}
		{\big\{ M (\rho_{j:j}-\rho_{:j} 
			\beta_{j}) \big\}}. \label{eq:exactS}
\end{align}
In the above formula the aspect errors are handled by using a
proper $P_{ij}(t)$ at each time bin and the non-uniformity is
handled by $\beta_j(t)$. The formula does not require an estimation of
the overall background level to remove the background pattern. One can
express the reconstructed image
$\hat{s}_j$ in terms of the true signal counts $s_j$ as
\begin{align}
	\frac{\{\hat{s}_j\}}{T} & = \frac{\big\{ s_j (\rho_{j:j}-\rho_{:j} 
			\beta_{j}) \big\}}
		{\big\{ \rho_{j:j}-\rho_{:j} 
			\beta_{j} \big\}}
	+\sum_{k \neq j} \frac{\big\{ s_k (\rho_{j:k}-\rho_{:k} 
			\beta_{j}) \big\}}
		{\big\{ \rho_{j:j}-\rho_{:j} 
			\beta_{j} \big\}}. 
\end{align} 
The second term in the above formula is the coding noise of the system.

\section{Refinements}

Here we study two refinement procedures of photon folding for suppressing
coding noise.  The first technique, recursive folding, is similar to
conventional recursive methods (e.g. IROS) \cite{Hammersley:92}, which
successively remove the side lobes -- coding noise of strong sources.
The second technique, high-order folding, represents a novel
approach to the coding-noise problem.

\subsection{Recursive Folding}

A simple way to remove coding noise is to photon fold
recursively.  For a given confidence $C$, the \ph\ of photon folding
will determine detections of signals in the image based on
Eq.~(\ref{eq:confi}).  Since the signal $\hat{s}_j$ is expected to
generate counts $\hat{h}_i$, we replace $h_i$ by
\begin{align}
	h_i & \leftarrow h_i-\hat{h}_i, \ \ \ \mbox{ where }\nonumber \\ 
	\hat{h}_i & = \sum_j
	\{\{\hat{s}_j\} P_{ij} \}/T, \end{align} and repeat the
procedure until no excess \ph\ is found for the given confidence.
The final image is the sum of $\hat{s}_j$ at all the recursion steps and
the residual.  This is very similar to many other recursive methods.  
Recursive folding is deterministic since it does not require an initial
guess and has a clear stopping point given by the confidence level.
This feature is missing in some image refinement procedures such as
MEM.

The above formula assumes that the source intensity and detector
background level are constant during the observation, or that the
detector orientation relative to the sky is fixed. In practice, this is
rarely the case. Without information on source intensity/background
history, the recursive method does not provide an effective correction
to coding noise.  In this sense, the recursive technique is not a
true solution for coding noise.  Now we study another approach to the
coding noise problem.

\subsection{High-Order Folding}

Consider the case where there is only one strong source ($s_l$) within
the FOV.  Given the detection of this source by regular photon
folding, one can again apply photon folding only in the detector space
shadowed by the mask from direction $l$ in the sky.  Since in the
shadowed region there is no contribution from source photons at
direction $l$, this second-order photon folding will provide an image
which is free of the coding noise from the source in direction~$l$.

We redefine the total counts and the total number of detector bins only in the
shadowed region for the given sky direction $l$ as following:
\begin{align}
	H_{l'}(t) & = H(t) \alpha_{l'}(t), \nonumber \\
	M_{l'}(t) & = \sum_{i:P_{il}(t)<\delta} 1.
\end{align} 

$\xi^2_{l'j}$ for the second-order photon folding will be given by 
\begin{align}
	\xi^2_{l'j}\Big|_{j\neq l} & = 
		\frac{\displaystyle 
		\big\{ H_{l'} \alpha_{l'j} - H_{l'} \beta_{l'j}\big\}^2}
		{\displaystyle \big\{ H_{l'} \beta_{l'j}\big\}}
		+
		\frac{\displaystyle 
		\big\{ H_{l'} \alpha_{l'j'} -  H_{l'} \beta_{l'j'}\big\}^2}
		{\displaystyle \big\{ H_{l'} \beta_{l'j'}\big\}} 
	\nonumber \\
	& = 
		\frac{\displaystyle \big\{ H_{l'}\big\} 
			\big\{ H_{l'} (\alpha_{l'j}
			- \beta_{l'j}) \big\}^2}
		{\displaystyle \big\{ H_{l'} \beta_{l'j}\big\}
			\big\{ H_{l'} (1-\beta_{l'j})\big\}}.
\end{align} 
The image by the second-order photon folding will be given as
\begin{align}
	\frac{\{\hat{s}_{l'j}\}_{j \neq l}}{T} & = \frac{\displaystyle
		\big\{ H_{l'} (\alpha_{l'j} - \beta_{l'j}) \big\}}
		{\big\{ M_{l'} (\rho_{l'j:j}-\rho_{l':j} 
			\beta_{l'j}) \big\}}. 
\end{align} 
The above two formulae are exactly the same as Eq.~(\ref{eq:exactXI}) and
(\ref{eq:exactS}) except for an additional index $l'$. The complete expression
for an arbitrary order of photon folding can be written by simply adding the
additional indices for each term (refer to Appendix).
In terms of true signal counts $s_j$, 
\begin{align}
	\frac{\{\hat{s}_{l'j}\}_{j \neq l}}{T} & = 
	\frac{\big\{ s_j (\rho_{l'j:j}-\rho_{l':j} 
			\beta_{l'j}) \big\}} 
		{\big\{ \rho_{l'j:j}-\rho_{l':j} 
			\beta_{l'j} \big\}} 
		+
	\sum_{k \neq j,l} \frac{\big\{ s_k (\rho_{l'j:k}-\rho_{l':k} 
			\beta_{l'j}) \big\}} 
		{\big\{ \rho_{l'j:j}-\rho_{l':j} 
			\beta_{l'j} \big\}} \nonumber \\
		& \ +
	\frac{\big\{ s_l (\rho_{l'j:l}-\rho_{l':l} 
			\beta_{l'j}) \big\}} 
		{\big\{ \rho_{l'j:j}-\rho_{l':j} 
			\beta_{l'j} \big\}}. \label{eq:sec}
\end{align} 

The second and third term in the above formula represent the coding
noise of the system from second-order photon folding. In general,
$\rho_{l'...:l}$ in the third term keeps the coding noise from the
strong source $s_l$ very small.  For example, in a system with $P_{ij}$
being either 0 or 1, there is no coding noise from $s_l$ since
$\rho_{l'...:l}=0$.  When there is a strong point source within the FOV,
the above method will provide the best image of weak sources within the
FOV regardless of the time dependence of the intensity of the strong
source relative to the background level.

There are a few problems in applying the above formula in practical
situations.  First, although the second-order folding removes the
coding noise from the primary strong sources, the coding noise from
other weak sources is larger in the second-order folding than in the
regular photon folding since the second-order folding utilizes only
part of the mask pattern or detector space. Such increase of coding
noise from weak sources does not guarantee the reduction of the overall
coding noises in high-order photon folding.

Second, the above formula might not be suitable for multiple strong
sources or an extended object.  In the presence of multiple strong
sources, one might have to rely on third or higher-order folding
involving the shadowed region for all the directions of strong
sources.  In typical coded aperture systems, $\rho \sim 0.5 $, so that
there is only $\rho^n$ fraction of the detector area for the completely
shadowed region from $n$ point sources.  That is, the shadowed region
for high-order folding runs out quickly as the number of sources
increases.

In order to overcome larger coding noise from weak sources in
second-order photon folding, the other region -- the unshadowed region
($\{\hat{s}_{lj}\}$) by the primary strong sources -- should be
utilized as well as the shadowed region ($\{\hat{s}_{l'j}\}$).  If the
detector background is uniform, the second-order photon folding in the
unshadowed region will provide similar results as in the shadowed
region. For example, in the case of the uniform background system with
$P_{ij}$ being either 0 or 1,  the term in $\{\hat{s}_{lj}\}_{j\ne l}$,
equivalent to the third term in Eq.~(\ref{eq:sec}), is proportional to
$\rho_{lj:l}-\rho_{l:l}\beta_{lj}$ which vanishes.

In general, the images by second-order folding, $\{\hat{s}_{lj}\}_{j\ne
l}$ and $\{\hat{s}_{l'j}\}_{j\ne l}$ will have substantially
reduced contributions from the coding noise of primary strong sources
at direction $l$. Both images, however, have larger coding noises
from the secondary sources.  The best second-order photon-folding
image will be a linear combination of these two images, i.e.
\begin{align} 
	\gamma \{\hat{s}_{l'j}\} + (1-\gamma) \{\hat{s}_{lj}\} \label{eq:weird}.
\end{align}

The optimal $\gamma$ strongly depends on $\delta$, $P_{ij}$, and etc.
For a given detection of a strong source by the regular photon
folding, the optimal $\gamma$ for the second-order photon folding can
be estimated by using only the strong source with constant intensity.
In other words, we first calculate $\{\hat{s}^0_{l'j}\}$,$\{\hat{s}^0_{lj}\}$ from
$\hat{h}_i$ instead of $h_i$, where
\begin{align}
\hat{h}_i & = \{\{\hat{s}_l\} P_{il} \}/T, \label{eq:feed}
\end{align}
and then use the $\gamma$ in Eq.~(\ref{eq:weird}), which minimizes
\begin{align} 
	\gamma \{\hat{s}^0_{l'j}\} + (1-\gamma) \{\hat{s}^0_{lj}\}.
\end{align} This procedure is valid when $s_l$ is much stronger than $s_j$,
which is the assumption of this technique.
Here once again we assume the constancy of the source
intensity in Eq.~(\ref{eq:feed}), but due to the terms like
$\rho_{l'...:l}$ in the coding noise from strong source $s_l$,
the changes of the source intensity are less serious in the
second-order folding than in the recursive method.

Now, we consider the case of multiple strong sources.
When $\{s_l\}\gg \{s_j\}$, the reconstructed images by the regular and
second-order folding can be summarized as
\begin{align}
	\{\hat{s}_j\} & = O(\tau_j) \{s_j\} + 
		\sum_{k \neq j} O(\tau_k\epsilon) \{s_k\},  \nonumber \\
	\gamma \{\hat{s}_{l'j}\} + (1-\gamma) \{\hat{s}_{lj} \} & = O(\tau_j)
		\{s_j\} + 
		\sum_{k \neq j,l} O(\tau_k\epsilon) \{s_k\} 
		+ O(\tau_l\epsilon^2) \{s_l\}, 
\end{align} where $\tau$ represents the temporal uniformity of the source
with respect to the background level and $\epsilon$ is the size of
coding noise.
Since the contributions of source $s_k|_{k\ne j,l}$ (the second term
in the right hand side of the above formula) should be similar
in both images from the first-order and second-order folding,
we expect that the following formula provides a more reliable estimate
for $s_j$ in the case of multiple strong sources.
\begin{align}
  	\sum_{l \neq j \atop \xi^2_l > \eta}\Big(\gamma \{\hat{s}_{l'j}\}  + (1-\gamma) \{\hat{s}_{lj}\} \Big) 
	& - (L-1) \{\hat{s}_j\} \nonumber \\
	&= O(\tau_j) \{s_j\} + \sum_{k \neq j} O(\tau_k \epsilon^2) \{s_k\}, \label{eq:final}
\end{align}  where $L$ is the number of sources detected in the
first-order photon folding and $\gamma$ is calculated by minimizing
\begin{align} 
	\sum_{l \neq j \atop \xi^2_l > \eta} \gamma \{\hat{s}^0_{l'j}\} + (1-\gamma) \{\hat{s}^0_{lj}\}.
\end{align} 
The left hand side of Eq.~(\ref{eq:final}) is expected to reduce the
coding noise below that from the first-order image and is less
sensitive to the intensity variation of the strong sources compared to the
recursive method.

\section{Simulations}

We demonstrate the photon folding technique on a coded-aperture
imaging system. The system consists of 64 $\times$ 64 random mask
elements ($P_{ij}$= 0 or 1) and the detector has 32 $\times$ 32
elements.

The first example has four point sources and one extended source within
the FOV.  Two of the point sources are very strong and the other sources
are relatively weak. During the simulated observation we assume that
the relative orientation of the mask pattern with respect to the sky has
rotated 90 degrees from the first half to the second half of the measurement.
This is a somewhat extreme case of the change of telescope orientation
relative to the sky during the observation.  When reconstructing
images, we combine all the data and we assume that the intensity change
of the sources during the observation is unknown, i.e.  the intensity
is assumed to be constant. The aspect system and spatial resolution of
the detector are assumed to be perfect.

\begin{figure} \begin{center} 
\epsfig{figure=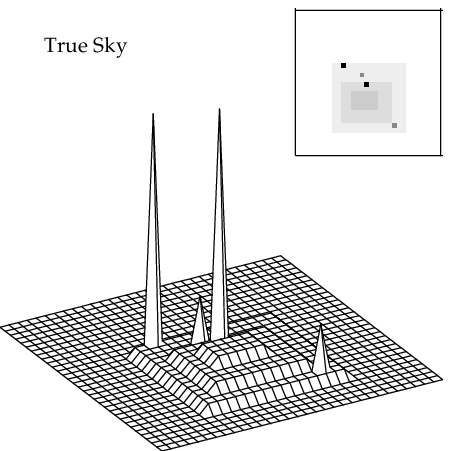,width=4.5cm,clip=} 
\epsfig{figure=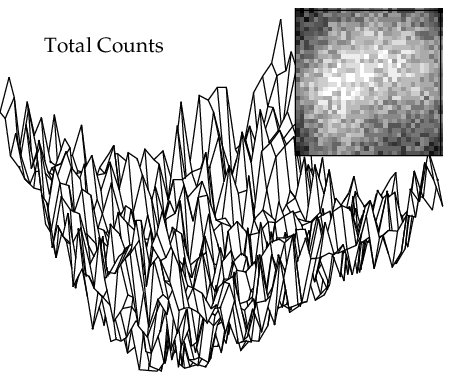,width=4.5cm,clip=}
\epsfig{figure=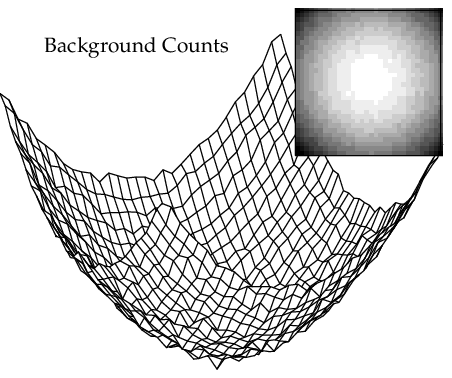,width=4.5cm,clip=}
\epsfig{figure=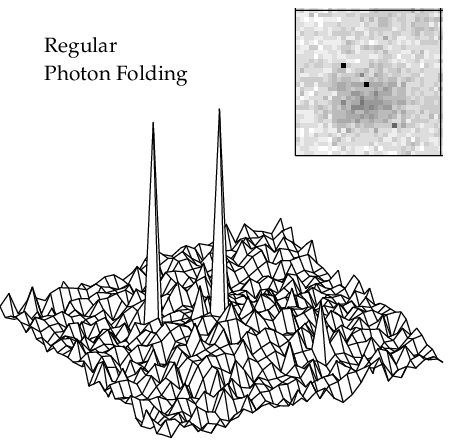,width=4.5cm,clip=}  
\epsfig{figure=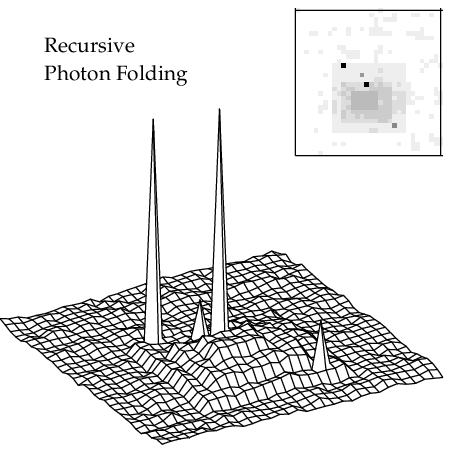,width=4.5cm,clip=}
\epsfig{figure=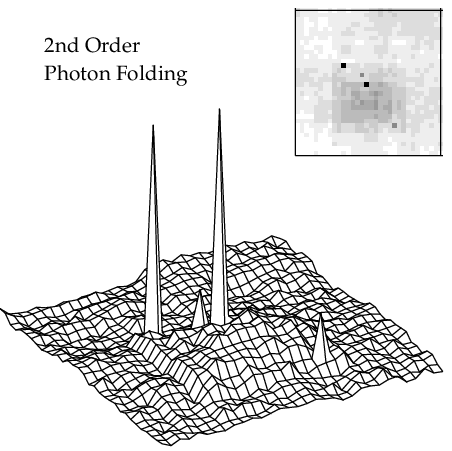,width=4.5cm,clip=}
\caption{Simulation of steady sources: the
upper plots are the simulated sky image, total counts, and background
counts in the detector. The lower plots show the reconstructed images by
regular photon folding, recursive  folding, and second-order folding.
The orientation of the telescope relative to the sky has changed 90
degrees from the first to the second half of the observation.}
\label{fig:fixed} \end{center} \end{figure}

The upper plots in Fig.~\ref{fig:fixed} show the true sky image and the
simulated detector counts.  We assume that the detector background
exhibits a quadratic spatial dependence, resulting in an enhancement of counts at
the edges. Such a pattern is somewhat common and we assume that the
pattern is known by previous flat field observations or calibrations.

In Fig.~\ref{fig:fixed}, the intensity of sources in the sky did not
vary during the simulated observation. The lower three plots show the
reconstructed images by regular photon folding, recursive folding, and
second-order folding.  The regular photon folding, which is equivalent
to the cross correlation method in this case, successfully detects the
strong sources, but the coding noise of the strong sources buries
features of weak sources.  The recursive folding presents the best
image in this case, and its quality is limited only by random Poisson
noise.  Second-order folding also reconstructs a decent image which
reveals the fine structure of weak sources.  The noise in the image by
second-order folding is mostly random Poisson noise and coding noise of
weak sources.

\begin{figure}[hb] \begin{center} 
\epsfig{figure=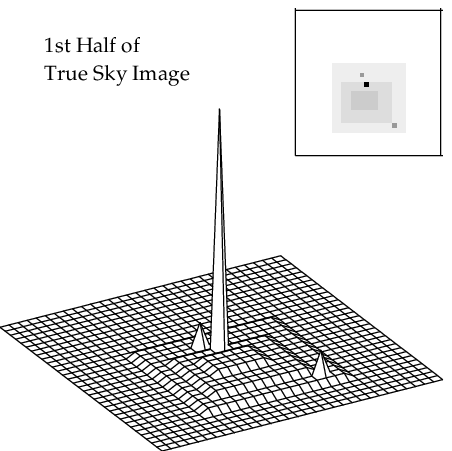,width=4.5cm,clip=}
\epsfig{figure=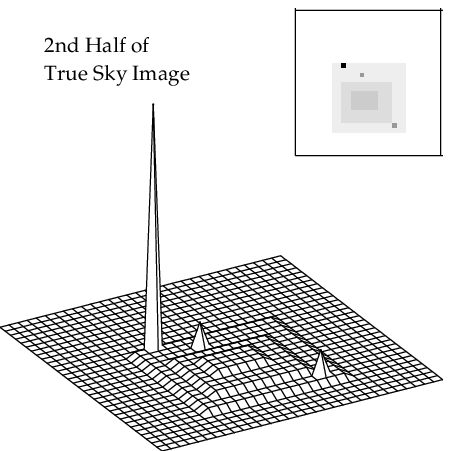,width=4.5cm,clip=} \\
\epsfig{figure=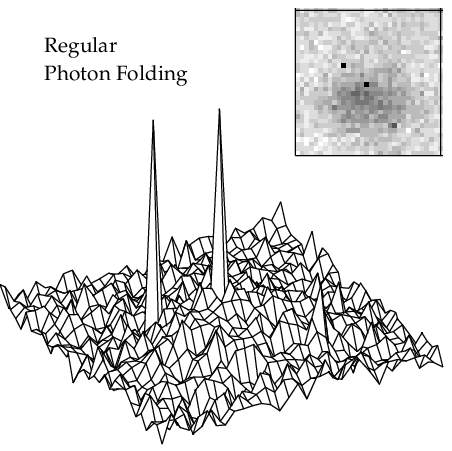,width=4.5cm,clip=}
\epsfig{figure=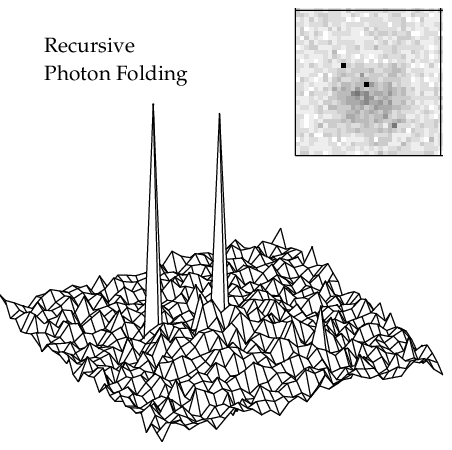,width=4.5cm,clip=}
\epsfig{figure=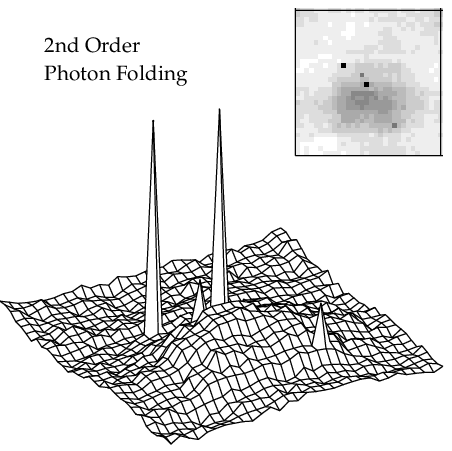,width=4.5cm,clip=} 
\caption{ Simulation of variable sources:
the upper plots show the true sky image in the first and second half of
the observation.  The orientation of the telescope relative to the sky
has changed 90 degrees from the first to the second half.  Reconstructed
images by the regular photon folding, recursive folding and second-order
photon folding are shown in the lower plots.} \label{fig:variable}
\end{center} \end{figure}

Now, in Fig.~\ref{fig:variable}, we have an extreme case of a more
realistic situation.  The intensity of two strong sources changed
dramatically from the first to the second half of the observation.  If the
relative orientation between the telescope and the sky did not change, the
results would be similar to those in Fig.~\ref{fig:fixed}. But here we
assume that the relative orientation changed 90 degrees.

Under the assumption of being unaware of the source intensity changes,
the bottom plots in Fig.~\ref{fig:variable} show the three
reconstructed images.  The recursive method fails to show the fine
structure of weak sources.  It should be noted that this result is
generally true for any type of recursive method without prior
information on source intensity changes.  On the other hand, the
second-order folding produces almost the same image as in the previous
case, i.e.~only limited by the random Poisson noise and coding noise of
weak sources

In order to see the general performance of the recursive method and the
second-order photon folding, we calculate SNRs of a
few simple simulations with a flat background pattern.  In the
following figure, we show the SNR of a point source in four distinct
situations (1 $\sigma$ distribution of the simulation results). Each
observation consists of two measurements as in the previous example (90
degree offset of the relative orientation).

\begin{figure} \begin{center} 
\epsfig{figure=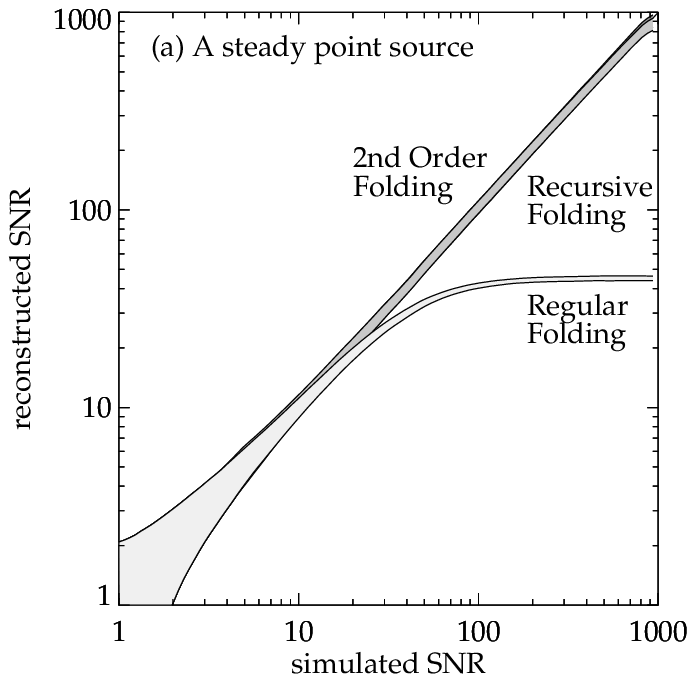,width=5.5cm,clip=}
\epsfig{figure=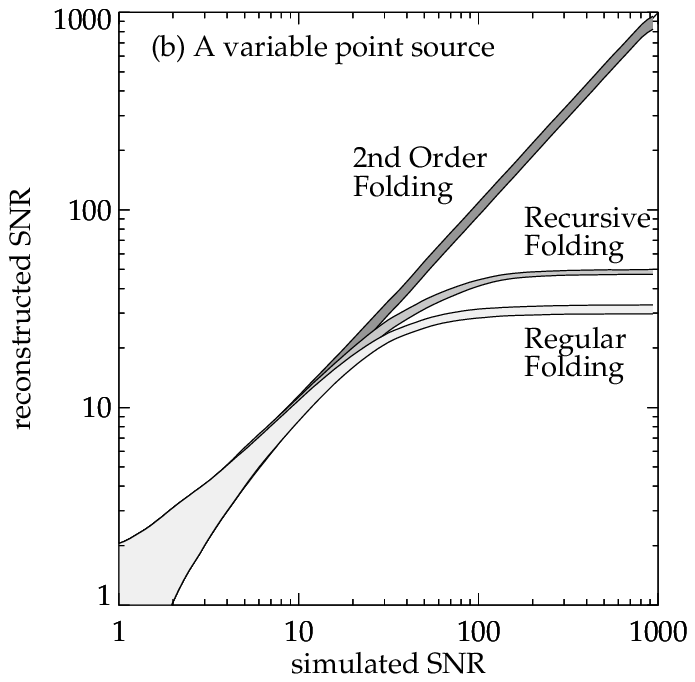,width=5.5cm,clip=}
\epsfig{figure=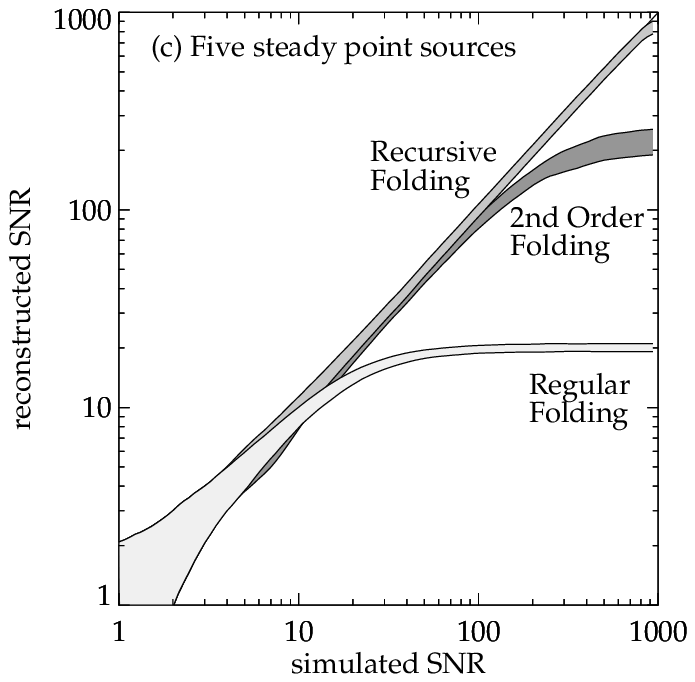,width=5.5cm,clip=}
\epsfig{figure=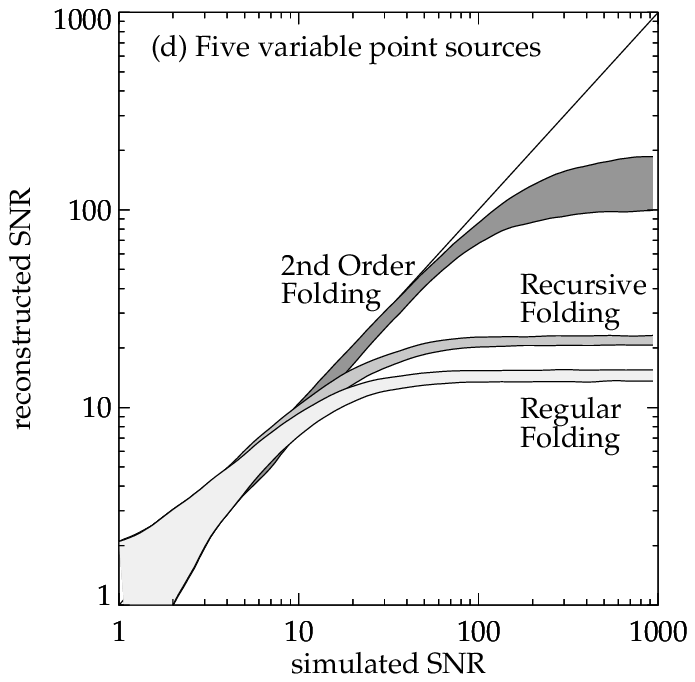,width=5.5cm,clip=}
\caption{SNRs in various cases: within the
FOV, there is (a) a steady point source (b) a variable point source, (c)
five steady point sources, and (d) five variable point sources.  The
overall intensity of each source in multiple-source cases are the same.}
\label{fig:snr} \end{center} \end{figure}

Fig.~\ref{fig:snr} (a) shows the case of a steady single source within
the FOV. The second-order folding and recursive folding generate images
with the maximally allowed SNR, while SNRs from regular folding are
limited by coding-noise.  In Fig.~\ref{fig:snr} (b),  the source
intensity dropped to zero for the second half.  While second-order
folding performs similarly to the previous cases, the performance of
recursive folding is limited by coding noises.

In Fig.~\ref{fig:snr} (c), there are
five steady point sources within the FOV. In this case, the recursive folding
produces perfect images, while second-order folding shows its limitation due
to incomplete correction of the coding noise from multiple sources.
In Fig.~\ref{fig:snr} (d), there are five variable point sources, and
the intensity of three sources drops from maximum to zero and the
intensity of the other two rises from zero to maximum from the first to the
second half of the observation.  It is clear that if there is a change
in source intensity, second-order folding is the better choice, while in
the case of many multiple steady sources recursive folding is the
optimal method.

\section{Discussion}

The photon folding and associated techniques can be applied in
many non-focusing instruments. Recursive folding and second-order
folding are complementary to each other in various realistic
situations.  Here we discuss some of the fundamental issues.

\subsection{Computational Issues}


Computational time for regular photon folding is similar to that of
the cross correlation method.  Since the photon folding automatically takes
care of aspect errors and non-uniform background, the overall
computational burden for photon folding is similar to regular
correlation or inversion techniques.

If we let $N$ be the number of calculations for photon folding, the
recursive folding requires $N_r N$ calculations where $N_r$ is the
total number of recursion steps. $N_r$ depends on intensity distribution of
sources in the FOV. Second-order folding requires $(2 N_{s} +1) N$
calculations, where $N_{s}$ is the number of strong sources detected in
the regular photon folding. The extra $N$ comes from the
estimation of the optimum $\gamma$ for the given detection. 
Since photon folding does not involve inversion of matrices, this 
number of calculations is not a problem even
for a huge number of sky or detector pixels due to the improvements of
modern computing power.

Both recursive and second-order folding can be implemented on-line.
While the recursive folding is somewhat straightforward in on-line
processing, second-order folding requires more caution. 
The successful image reconstruction of second-order folding depends on
limiting the number ($L$) of sky pixels for the first-order detection
and yet including enough of them to remove most of the coding
noise. Selecting a few of the strongest sources for the first detection
would be adequate for an automatic implementation.

\subsection{Recursive vs Second-order folding}

\begin{figure} \begin{center} 
\epsfig{figure=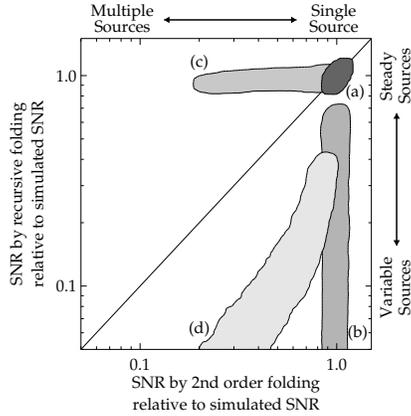,width=5.5cm,clip=}
\caption{Recursive vs Second-order photon
folding:  (a) a steady source (b) a variable source, (c) five steady
sources, and (d) five variable sources.  The $x$-axis is the SNR by
second-order photon folding relative to the simulated SNR, and $y$-axis
the SNR by recursive folding folding relative to the simulated SNR.
Each shaded region represents each of the four cases of simulations in
Fig.~\ref{fig:snr} where simulated SNR is between 50 and 1000.}
\label{fig:com} \end{center} \end{figure}

These two techniques are somewhat complementary to each other in
various situations. Fig.~\ref{fig:com} summarizes the simulations in
the previous section (Fig.~\ref{fig:snr}) in terms of relative SNR.  In
Fig.~\ref{fig:com}, each of the shaded regions represents each
case of the simulations in Fig.~\ref{fig:snr} where simulated SNR is 
between 50 and 1000.  For a given value of the simulated SNR, recursive
folding produces better results in the case of many steady sources,
while second-order folding does so in the case of variable sources.

The second-order photon folding is less robust against aspect errors or
non-uniformity of the background than the recursive folding.  Having
uncorrectable aspect errors is somewhat equivalent to having multiple
sources in the FOV.  The non-uniform background in the second-order
image ($\{\hat{s}_{lj}\}$) from the unshadowed region cannot cancel out
in photon folding, depending on the strength of the strong source $s_l$
relative to the background level.

\subsection{Other variations of photon folding}

So far we use only two folded bins -- shadowed/unshadowed region -- for
photon folding.  In general, increasing the number of folded bins does
not boost SNR.  However it may be useful to have more than two folded
bins when effective transparencies have more than two distinct
values or there are severe non-uniformities in the detector background.
One can also utilize higher-order folding beyond second-order, such as
$\hat{s}_{l'k'j}$, $\hat{s}_{l'kj}$, etc.  Although higher-order
foldings are computationally intensive, the use of higher-order
folding could remove the limitation of the second-order folding method in multiple
strong source cases.  Further studies are required to find a proper
form of higher-order folding to suppress even more of the coding noise and
also new types of optics scheme can be designed to optimize for
high-order photon-folding image reconstruction.


\subsection{Applications}

The photon folding technique is very versatile, and it can be applied
to many different types of experiments.  Applying the photon-folding
method to other non-focusing systems like Fourier grid systems is quite
straightforward. In the case of modulation collimator systems, the
location of the source is identified by the temporal modulation rather
than the spatial modulation in the detector (usually detectors for
modulation collimator systems do not require spatial resolution).

For example, consider a typical rotation modulation collimator.  For a
given direction in the sky, the fraction ($\rho$) of the shadowed area
in the detector changes with time. For a given unit rotation and a
given direction in the sky, one can rearrange count rates in the order
of the fraction ($\rho$) instead of time, and then apply a folding
procedure with a necessary amount of binning. This is somewhat similar
to regular epoch-folding in timing analysis.  The signal from a
point source usually fluctuates between a maximum and minimum during
the unit rotation.  The temporal resolution, relatively finer than the
spatial resolution of a typical detector, might allow effective
usage of more than two folded-bins in photon folding.


In coded-aperture systems, photon folding may allow use of non-URA mask
patterns.  For example, {\it EXIST} is a coded-aperture, wide field of
view survey mission, with a wide energy range ($\sim$ 10 -- 600 keV)
\cite{exist:99}.  Curved mask patterns are being studied for {\it EXIST}
to utilize maximal available FOV without substantial collimators between
the mask and detectors.  In order to cover the wide energy range (10 --
600 keV) in {\it EXIST}, two scale mask patterns can be used with energy
dependent transparency \cite{Skinner:93}.  Such mask patterns may
provide additional advantages over conventional mask patterns, but the
optimal imaging reconstruction scheme is not yet available.  The photon
folding technique may be useful for non-conventional mask patterns.
Particularly second-order folding is very interesting for {\it EXIST}.
The high sensitivity of {\it EXIST} is likely to result in detection of
multiple, possibly variable sources in each field, and second-order
folding would allow imaging of the weak sources.

\section{Conclusion}

A new imaging technique -- photon folding -- is introduced  for
non-focusing telescopes. The technique is quite robust against common
imaging problems like aspect errors and non-uniform background.  Its
performance is demonstrated by a two-dimensional coded-aperture system
and photon folding can be applied to other types of imaging
telescopes.  Two refinements of photon folding are presented
-- recursive and second-order folding.  In particular second-order
photon folding is suitable for imaging weak sources in the presence of
highly variable strong sources regardless of the changes in telescope
orientation relative to the sky.

\section{Acknowledgement}

The author would like to thank C.J. Hailey for valuable comments and
discussion.

\appendix

\section*{Appendix A}

For the regular photon folding,
\begin{align}
	\alpha_j(t) & = \frac{\lambda_{j:}[h_i(t),t]}{\lambda_{:}[h_i(t),t]} 
 		 = \frac{1}{H(t)} \sum_{i:P_{ij}(t)\ge\delta} h_i(t),
\end{align} which reduces to Eq.~(\ref{eq:need}) when $P_{ij}$ is
either 0 or 1 (constant).
\begin{align}
	\beta_j(t) & = \frac{\lambda_{j:}[\bar{b}_i(t),t]}{\lambda_{:}[\bar{b}_i(t),t]} 
 		 = \frac{1}{\bar{B}(t)} \sum_{i:P_{ij}(t)\ge\delta} \bar{b}_i(t),
\end{align} where $\bar{B}(t)=\Sigma \bar{b}_i(t)$.
For a flat background pattern, $\beta_j(t) = \rho_{j:}(t)$.

And
\begin{align}
	\rho_{j:}(t) & = \frac{\lambda_{j:}[1,t]}{\lambda_{:}[1,t]} 
		 = \frac{1}{M(t)} \sum_{i:P_{ij}(t)\ge\delta} 1, \nonumber \\
	\rho_{j:j}(t) & = \frac{\lambda_{j:j}[1,t]}{\lambda_{:}[1,t]} 
		  = \frac{1}{\displaystyle M(t)} 
		\sum_{i:P_{ij}(t)\ge\delta} P_{ij}(t), \nonumber \\
	\rho_{:j}(t) & = \frac{\lambda_{:j}[1,t]}{1}  
		= \frac{1}{M(t)} \sum_{i} P_{ij}(t).
\end{align}
If $P_{ij}(t)$ is either 0 or 1, $\rho_{:j}(t)=\rho_{j:j}(t)=\rho_{j:}(t)$.

Now, for second-order photon folding,
\begin{align}
	\alpha_{lj}(t) & = \frac{\lambda_{lj:}[h_i(t),t]}{\lambda_{l:}[h_i(t),t]} \nonumber  \\
 		 & = \Bigg[\frac{\displaystyle 1}{H(t)} 
			\sum_{i:P_{il}(t)\ge\delta, \atop P_{ij}(t)\ge\delta} h_i(t) \Bigg]
 		 \Bigg[ \displaystyle \frac{1}{H(t)} 
			\sum_{i:P_{il}(t)\ge\delta} h_i(t) \Bigg]^{-1}  \nonumber\\
		& = \frac{\displaystyle 1}{\displaystyle H(t) \alpha_l(t)}
		\sum_{i:P_{il}(t)\ge\delta, \atop P_{ij}(t)\ge\delta} h_i(t) 
		 = \frac{1}{H_l(t)} \sum_{i:P_{il}(t)\ge\delta, \atop P_{ij}(t)\ge\delta} h_i(t).
\end{align}  Likewise, 
\begin{align}
	\beta_{lj}(t) & = \frac{1}{\bar{B}_l(t)} 
		\sum_{i:P_{il}(t)\ge\delta, \atop P_{ij}(t)\ge\delta} \bar{b}_i(t) \nonumber \\
	\rho_{lj:}(t) & = \frac{1}{M_l(t)} 
		\sum_{i:P_{il}(t)\ge\delta, \atop P_{ij}(t)\ge\delta} 1,  \nonumber \\
	\rho_{lj:j}(t) & = \frac{1}{\displaystyle M_l(t)} 
		\sum_{i:P_{il}(t)\ge\delta, \atop P_{ij}(t)\ge\delta} P_{ij}(t), \nonumber \\
	\rho_{l:j}(t) & = \frac{1}{M_l(t)} \sum_{i:P_{il}(t)\ge\delta} P_{ij}(t),
\end{align} where
\begin{align}
 	\bar{B}_l(t) & = \bar{B} (t) \beta_l(t), \nonumber \\
	M_{l}(t) & = \sum_{i:P_{il}(t)\ge\delta} 1.
\end{align} 
In $\alpha_{l'j}(t)$, $\beta_{l'j}(t)$, $\rho_{l'j:}(t)$,
$\rho_{l'j:j}(t)$ and $\rho_{l':j}(t)$, the summation condition,
$P_{il}(t)\ge\delta$, in the above equations will be replaced by
$P_{il}(t)<\delta$. Therefore, they satisfy the followings relations.
\begin{align}
	\alpha_{lj}(t) + \alpha_{lj'}(t) & = 1, \nonumber \\
	\beta_{lj}(t) + \beta_{lj'}(t) & = 1.
\end{align}

The complete expression for
an arbitrary order of photon folding can be written simply by keeping
additional indices in front of each term.
\begin{align}
	\xi^2_{l_1...l_nj}\Big|_{j\neq l_1...l_n} 
	& = 
		\frac{\displaystyle \big\{ H_{l_1...l_n}\big\} 
			\big\{ H_{l_1...l_n} (\alpha_{l_1...l_nj}
			- \beta_{l_1...l_nj}) \big\}^2}
		{\displaystyle \big\{ H_{l_1...l_n} \beta_{l_1...l_nj}\big\}
			\big\{ H_{l_1...l_n} (1-\beta_{l_1...l_nj})\big\}}, \nonumber \\
	\frac{\{\hat{s}_{l_1...l_nj}\}_{j \neq l_1...l_n}}{T} & =
		\frac{\displaystyle \big\{ H_{l_1...l_n} (\alpha_{l_1...l_nj} -
			\beta_{l_1...l_nj}) \big\}} {\big\{ M_{l_1...l_n}
			(\rho_{l_1...l_nj:j}-\rho_{l_1...l_n:j} \beta_{l_1...l_nj}) \big\}}. 
\end{align}


\begin{thebibliography}{10}

\bibitem{Fenimore:78}
	E.~E.~{Fenimore},
	``Coded aperture imaging: predicted performance of uniformly redundant
	arrays,"
	Appl.~Opt. {\bf 17}, 3562--3570 (1978).

\bibitem{Heft:97}
	C.~J.~{Hailey}, S.~{Abdali}, F.~E.~{Christensen}, W.~W.~{Craig}, 
		T.~R.~{Decker}, F.~A.~{Harrison}, and M.~{Jimenez-Garate},
	``Substrates and mounting techniques for the High-Energy Focusing
	  	Telescope (HEFT),"
	In {\em Proc. SPIE 3114: EUV, X-Ray, and Gamma-Ray Instrumentation
	  	for Astronomy VIII} 
	(International Society for Optical Engineering, Bellingham
		Washington, 1997), 
	Vol.~3114, pp.~535--543.

\bibitem{constX:00}
	N.~E.~{White} and H.~{Tananbaum},
	``The Constellation X-ray Mission,"
	In {\em IAU Symp.~195: Highly Energetic Physical Processes and
	  	Mechanisms for Emission from Astrophysical Plasmas}, 
	(Astronomical Society of the Pacific, San Francisco, California, 2000),
	Vol.~195, pp.~61--68.

\bibitem{exist:99}
	J.~E.~{Grindlay}, L.~{Bildsten}, D.~{Chakrabarty}, M.~{Elvis},
		A.~{Fabian}, F.~{Fiore}, N.~{Gehrels}, C.~{Hailey}, F.~{Harrison},
		D.~{Hartmann}, T.~{Prince}, B.~{Ramsey}, R.~{Rothschild},
		G.~{Skinner}, and S.~{Woosley},
	``EXIST: A High Sensitivity Hard X-ray Imaging Sky Survey Mission for
		ISS,"
	In {\em AIP conference proceedings: The Fifth Compton Symposium}
   	(American Institute Of Physics, Melville, New York, 1999), 
	Vol.~510, pp.~784--788.

\bibitem{BeppoSax:99}
	F.~A.~{Harrison}, J.~S.~{Bloom}, D.~A.~{Frail}, R.~{Sari}, 
		S.~R.~{Kulkarni}, S.~G. {Djorgovski}, T.~{Axelrod}, J.~{Mould},
		B.~P.~{Schmidt}, M.~H.~{Wieringa}, R.~M.~{Wark}, R.~{Subrahmanyan},
		D.~{McConnell}, P.~J.~{McCarthy}, B.~E.~{Schaefer},
		R.~G.~{McMahon}, R.~O.~{Markze}, E.~{Firth}, P.~{Soffitta}, and
		L.~{Amati},
	``Optical and Radio Observations of the Afterglow from GRB 990510:
		  Evidence for a Jet,"
	Astrophys. J. {\bf 523}, L121--L124 (1999).

\bibitem{Swift:99}
	N.~{Gehrels} and {Swift Science Team},
	``Swift - The Next GRB MIDEX Mission,"
	In {\em American Astronomical Society Meeting}, 
	(American Astronomical Society, Washington, DC, Dec, 1999), 
	Vol.~195, \#92.08.

\bibitem{Pixon:99}
	R.~C. {Puetter} and A.~{Yahil},
	``The Pixon Method of Image Reconstruction,"
	In {\em ASP Conf. Ser. 172: Astronomical Data Analysis Software and
	  	Systems VIII} 
	(Astronomical Society of the Pacific, San Francisco, California,
		1999), 
	Vol.~8, pp.~307--316.

\bibitem{MemPixon:97}
	T.~R. {Metcalf}, D.~{Alexander}, N.~{Nitta}, and T.~{Kosugi},
	``A Comparison of the MEM and Pixon Algorithms for HXT Image
	  	Reconstruction,"
	In {\em American Astronomical Society, Solar Physics Division Meeting}
	(American Astronomical Society, Washington, DC, May, 1997),
 	Vol.~28, \#02.17.

\bibitem{epf:83}
	D.~A.~{Leahy}, W.~{Darbro}, R.~F.~{Elsner}, M.~C.~{Weisskopf},
		S.~{Kahn}, P.~G.~{Sutherland}, and J.~E.~{Grindlay}.
	``On searches for pulsed emission with application to four globular
	  	cluster X-ray sources - NGC 1851, 6441, 6624, and 6712,"
	Astrophys.~J.~{\bf 266}, 160--170 (1983).

\bibitem{epf:96}
	S.~{Larsson},
	``Parameter estimation in epoch folding analysis,"
	Astro.~Astrophys.~Suppl.~{\bf 117}, 197--201 (1996).

\bibitem{Hammersley:92}
	A.~Hammersley, T.~Ponman, and G.~Skinner,
	``Reconstruction of images from a coded-aperture box camera,"
	Nucl.~Instr.~Meth.~{\bf A311}, 585--594 (1992).

\bibitem{Skinner:95}
	{G.~K.~Skinner, R.~L.~Balthazor, J.~R.~H.~Herring, R.~M.~Rideout,
		J.~Tueller, S.~D.~Barthelmy and L.~M.~Bartlett},
	``A balloon flight test of a coded-mask telescope with a multi-element
		  germanium detector,"
	Nucl. Instr. Meth.~{\bf A357}, 580--587 (1995).

\bibitem{Skinner:93}
	G.~K.~{Skinner} and J.~E.~{Grindlay},
	``Coded masks with two spatial scales,"
	Astron.~Astrophys.~{\bf 276}, 673--681 (1993).

\end{thebibliography}

\end{document}